\documentclass[aps,prc,preprint,superscriptaddress,showpacs,eqsecnum]{revtex4}

\usepackage{graphicx}
\usepackage{dcolumn}
\usepackage{bm}

\newcommand{\Sna}{$^{112}$Sn}
\newcommand{\Snb}{$^{124}$Sn}
\newcommand{\Snc}{$^{112,124}$Sn}
\newcommand{\raa}{($\alpha$,$\alpha$)}
\newcommand{\rgn}{($\gamma$,n)}

\newcommand{\rga}{($\gamma$,$\alpha$)}
\newcommand{\rag}{($\alpha$,$\gamma$)}
\newcommand{\ran}{($\alpha$,n)}
\newcommand{\rgp}{($\gamma$,p)}

\newcommand{\anucpot}{$\alpha$--nucleus potential}
\newcommand{\anucpots}{$\alpha$--nucleus potentials}

\begin{document}

\title{
Elastic $\alpha$-scattering of $^{112}$Sn and $^{124}$Sn at
astrophysically relevant energies
}

\author{D.\ Galaviz}
\email{galaviz@nscl.msu.edu}
\altaffiliation{Present address: NSCL, Michigan State University, 1
  Cyclotron Lab, East Lansing, MI 48824-1321, USA
}
\affiliation{
Institut f\"ur Kernphysik, Technische Universit\"at Darmstadt,
Schlossgartenstra{\ss}e 9, D-64289 Darmstadt, Germany
}
\author{Zs.~F\"ul\"op}
\author{Gy.\ Gy\"urky}
\author{Z.~M\'at\'e}
\affiliation{
	ATOMKI, PO Box 51, H-4001 Debrecen, Hungary
}
\author{P.\ Mohr}
\altaffiliation{Present address: Strahlentherapie, Diakoniekrankenhaus
  Schw\"abisch Hall, D-74523 Schw\"abisch Hall, Germany} 
\affiliation{
Institut f\"ur Kernphysik, Technische Universit\"at Darmstadt,
Schlossgartenstra{\ss}e 9, D-64289 Darmstadt, Germany
}
\author{T.~Rauscher}
\affiliation{
	Departement f\"ur Physik und Astronomie, Universit\"at Basel, 
	Klingelbergstrasse 82, CH-4056 Basel, Switzerland
}
\author{E.~Somorjai}
\affiliation{
	ATOMKI, PO Box 51, H-4001 Debrecen, Hungary
}
\author{A.\ Zilges}
\affiliation{
Institut f\"ur Kernphysik, Technische Universit\"at Darmstadt,
Schlossgartenstra{\ss}e 9, D-64289 Darmstadt, Germany
}
\date{\today}
\begin{abstract}
The cross sections for the elastic scattering reactions \Snc \raa \Snc\
at energies above and below the Coulomb barrier are presented and
compared to predictions for global \anucpots. The high precision of
the new data allows a study of the global \anucpots\ at both the proton
and neutron-rich sides of an isotopic chain. In addition, local
\anucpots\ have been extracted for both nuclei, and used to
reproduce elastic scattering data at higher energies. Predictions from
the capture cross section of the reaction \Sna \rag $^{116}$Te at
astrophysically relevant energies are presented and compared to
experimental data.

\end{abstract}

\pacs{24.10.Ht, 25.55.-e, 25.55.Ci, 26.30.+k}


\maketitle

\section{Introduction}
\label{sec:intro}
The 35 most proton-rich, stable isotopes between Se and Hg are called
{\it p} nuclei. Contrary to the synthesis of most nuclei above Fe through
neutron captures in the {\it s} and {\it r} process \cite{Burb57}, the
production of the {\it p} nuclei proceeds mainly via photon--induced
reactions at temperatures around a few GK \cite{Woos78,Lang01}. Seed nuclei 
already present in the stellar plasma and originating from the {\it s} and
{\it r} process are disintegrated mainly by \rgn, \rgp\ and \rga\
reactions in the thermal photon bath of the corresponding explosive
astrophysical event. Due to persisting problems in the reproduction of
the observed {\it p} abundances, a definite conclusion on the actual
site of this nucleosynthesis process -- usually called {\it p} or
$\gamma$ process -- cannot be drawn yet, however, the commonly
most favored site providing the required conditions is explosive Ne/O burning
in type II supernovae \cite{Woos78,Raus02,Arno03}. Recently,
consistent studies of {\it p} nucleosynthesis have become available,
employing theoretical reaction rates in large reaction
networks \cite{Arno03,Raus02,Cost00}. For heavy nuclei (140 $\lesssim
A \lesssim$ 200), \rgn\ and \rga\ reactions play the dominant role while other
photon-induced reactions are practically negligible. This is not the
case for lighter nuclei where captures and photodisintegrations
involving neutrons, protons, and $\alpha$ particles have to be
considered \cite{Rapp04,Raus05}. Finally, neutrino-induced reactions
may have some importance for selected isotopes because of the high
neutrino-flux stemming from the core collapse triggering the type II
supernova explosion \cite{Arno03,Raus02}. 

The cross sections used to calculate the required astrophysical reaction
rates in network studies are based on the statistical model
(Hauser-Feshbach) \cite{Haus52,Raus98,Gori00,Raus01,Raus04}. Global optical
potentials are considered in these calculations encompassing many
hundreds of nuclei and several thousand reactions. Experimental data
is scarce due to the sub-Coulomb energies and the large number of
unstable nuclei relevant for astrophysical applications. Only
recently, a number of experiments has been devoted to the study of
cross sections at astrophysically relevant energies. However, there is
still a lack of relevant experimental data for \rga\ and \rag\
reactions because of the high Coulomb barriers. Recent
$\alpha$ capture experiments on heavy nuclei at astrophysically relevant
energies were performed on $^{70}$Ge \cite{Fulo96}, $^{96}$Ru
\cite{Rapp02}, \Sna\ \cite{Ozka02} and $^{144}$Sm \cite{Somo98}. No
experimental data for \rga\ cross sections are available at
astrophysically interesting energies.

In general, \rag\ and \rga\ reaction cross sections are very sensitive
to the choice of the \anucpot, as has been observed in the
huge uncertainties for the prediction of the $^{144}$Sm\rag $^{148}$Gd cross
section \cite{Somo98,Mohr97,Raus95}. Elastic $\alpha$ scattering at
low energies (close to the Coulomb barrier) should provide an
additional test for the \anucpots\ considered in {\it p} process
network calculations. High precision data are needed for a clear
determination of the optical potential properties at the measured energies.

In this work the cross sections for the reactions \Snc \raa \Snc\ at
energies above and below the Coulomb barrier are presented. The new
experimental data provide a test for the global parameterizations
considered in {\it p} process network calculations. Furthermore, the
study of both proton- and neutron-rich stable tin isotopes provides
important information about the variation of \anucpots\ along an
isotopic chain. A local \anucpot\ is derived for both
neutron--deficient (\Sna) and neutron--rich (\Snb) nuclei.

In this paper we first present the experimental procedure
(Sec.~\ref{sec:exp_proc}) and compare the results to existing global \anucpots\
(Sec.~\ref{subsec:ang_theo}). The determination of the potential for
the tin isotopes is performed within the Optical Model (OM) framework
(Sec.~\ref{subsec:ang_pres}), and the results are compared to previous
experimental data (Secs.~\ref{subsec:ang_dist}--D). In
addition, the derived potential for \Sna\ is used for a prediction
of the \Sna \rag $^{116}$Te cross section which has been recently
measured using the activation technique \cite{Ozka02}
(Sec.~\ref{sec:sn_ag_te}).

\section{Experimental procedure and results}
\label{sec:exp_proc}
The scattering experiments were performed at the cyclotron laboratory of
ATOMKI, Debrecen, Hungary, where $^4$He$^{2+}$ beams are available up
to energies of about $E_\alpha$\,=\,20\,MeV. Angular distributions
were measured for $^{112}$Sn at $E_\alpha$\,=\,19.5\,MeV and 14.4\,MeV,
and for $^{124}$Sn at $E_\alpha$\,=\,19.5\,MeV. The beam intensity was
approximately 300~nA. The
experimental setup was similar to the one used in our previous
experiments on $^{144}$Sm \cite{Mohr97} and $^{92}$Mo
\cite{Fulo01}. Further experimental details on the setup can be found in
\cite{Mate89}. Here we briefly summarize the most important features
of the setup.

The angular distributions were measured using four silicon
surface-barrier detectors mounted on two turntables. The
solid angles varied between $1.55 \times 10^{-4}$\,sr and $1.63 \times
10^{-4}$\,sr. Two additional detectors placed at $15^\circ$ left
and right to the incoming beam axis with solid angles of $8.1 \times
10^{-6}$\,sr are used to normalize the angular distributions and to
determine the beam position on the target with high precision. Note that small
horizontal deviations of about 1\,mm lead to corrections of the
cross section of the order of 1\,\% at very forward angles. These
corrections can be precisely determined from the ratio of the count
rates in the two monitor detectors.

The targets consisted of highly enriched $^{112}$Sn (99.6\,\%) and
$^{124}$Sn (97.4\,\%) deposited onto thin carbon backings. The target
thicknesses of 200\,$\mu$g/cm$^2$ for both isotopes were confirmed by
the measured Rutherford cross sections during the experiment. The
absolute normalization was obtained from the Rutherford cross section
at very forward angles~\cite{Mohr97,Fulo01}.

\begin{figure}[hbt]
\includegraphics[ width = 85 mm, clip]{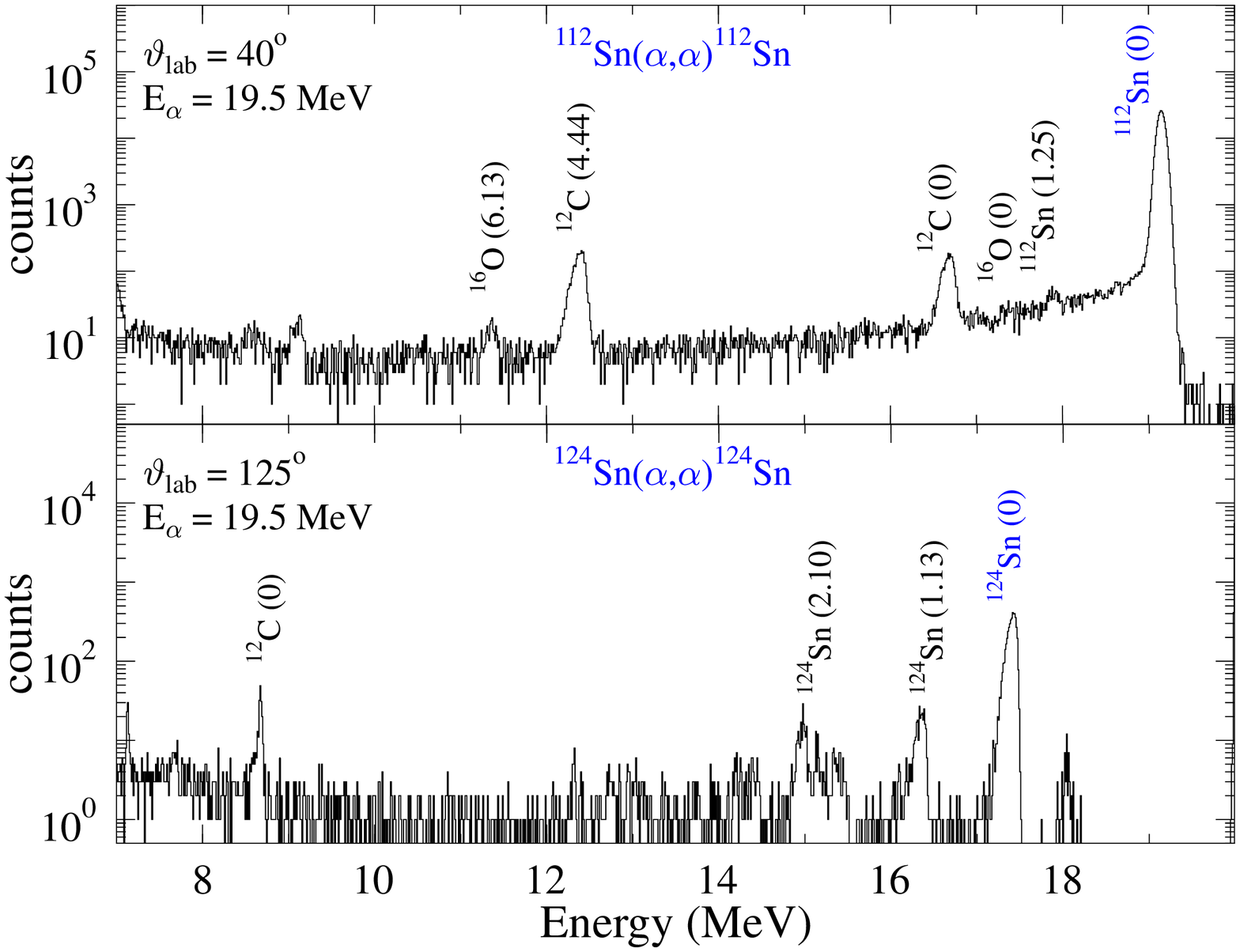}
\caption{
\label{fig:spec} 
Spectra of the \Sna \raa \Sna\ and \Snb \raa \Snb\ reactions at
$E_\alpha$\,=\,19.5\,MeV at forward ($\vartheta_{\rm{lab}} = 40^\circ$)
and backward ($\vartheta_{\rm{lab}} = 125^\circ$) angles,
respectively. Elastic scattering on target contaminations (mainly
$^{12}$C from the carbon backing) and inelastically scattered events
are clearly separated from the elastic peak. A small target
contamination with $A \approx 200$ which is visible
only in the spectra at backward angles remains below 1\,\%
contribution to the elastic peaks. The pulser signals used for dead
time correction are not seen in the spectra, as they lay above the
shown energy region.
}
\end{figure}

A precise dead time correction is necessary especially at forward angles
where the cross sections are huge. Therefore, the dead
time was monitored using pulse generators in all spectra. Typical
spectra of the \Sna \raa \Sna\ and \Snb \raa \Snb\ reactions are shown
in Fig.~\ref{fig:spec}. 

\begin{figure}[hbt]
\includegraphics[ width = 85 mm, clip]{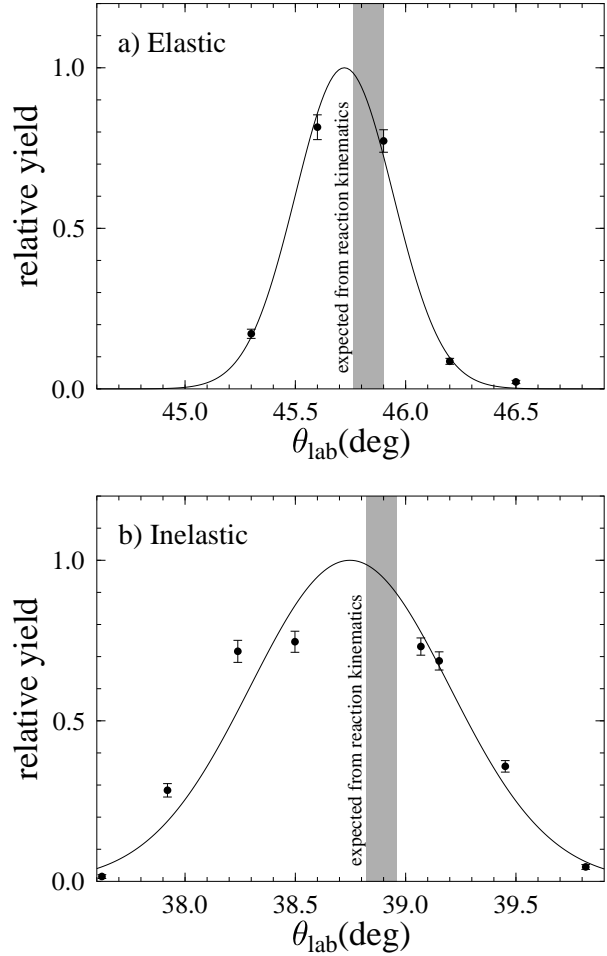}
\caption{
\label{fig:ang_calib}
Relative yield of $^{12}$C recoil nuclei in coincidence with
a) elastically and b) inelastically scattered $\alpha$ particles. A
Gaussian fit to the experimental data (solid line) is shown to guide
the eye. The shaded area presents the angle and the uncertainty
expected from the measured reaction kinematics.
}
\end{figure}

From a kinematic coincidence between the $\alpha$ particles scattered
on a 20 $\mu$g/cm$^2$ carbon backing foil and the $^{12}$C recoil nuclei, we
calibrated the position of the silicon detectors with a precision of
$\Delta \vartheta$ = 0.07$^\circ$. The scattered $\alpha$ particles
were measured using one detector placed at $\vartheta_{lab}$ =
70$^\circ$ (right side relative to beam axis). The recoil $^{12}$C
nuclei from the elastic ($^{12}$C$_{gs}$) and inelastic ($^{12}$C$_{2^+}$,
E$_x$=4.44~MeV) scattering were measured with another detector (left
side) which was moved around the expected positions,
$\vartheta_{lab,el.}$ = 45.83$^\circ$ and
$\vartheta_{lab,inel.}$ = 38.89$^\circ$. The results are shown in
Fig.~\ref{fig:ang_calib}. The maximum recoil rate is observed at the
expected position within the statistical uncertainties. 

\begin{figure}[hbt]
\includegraphics[ width = 85 mm, clip]{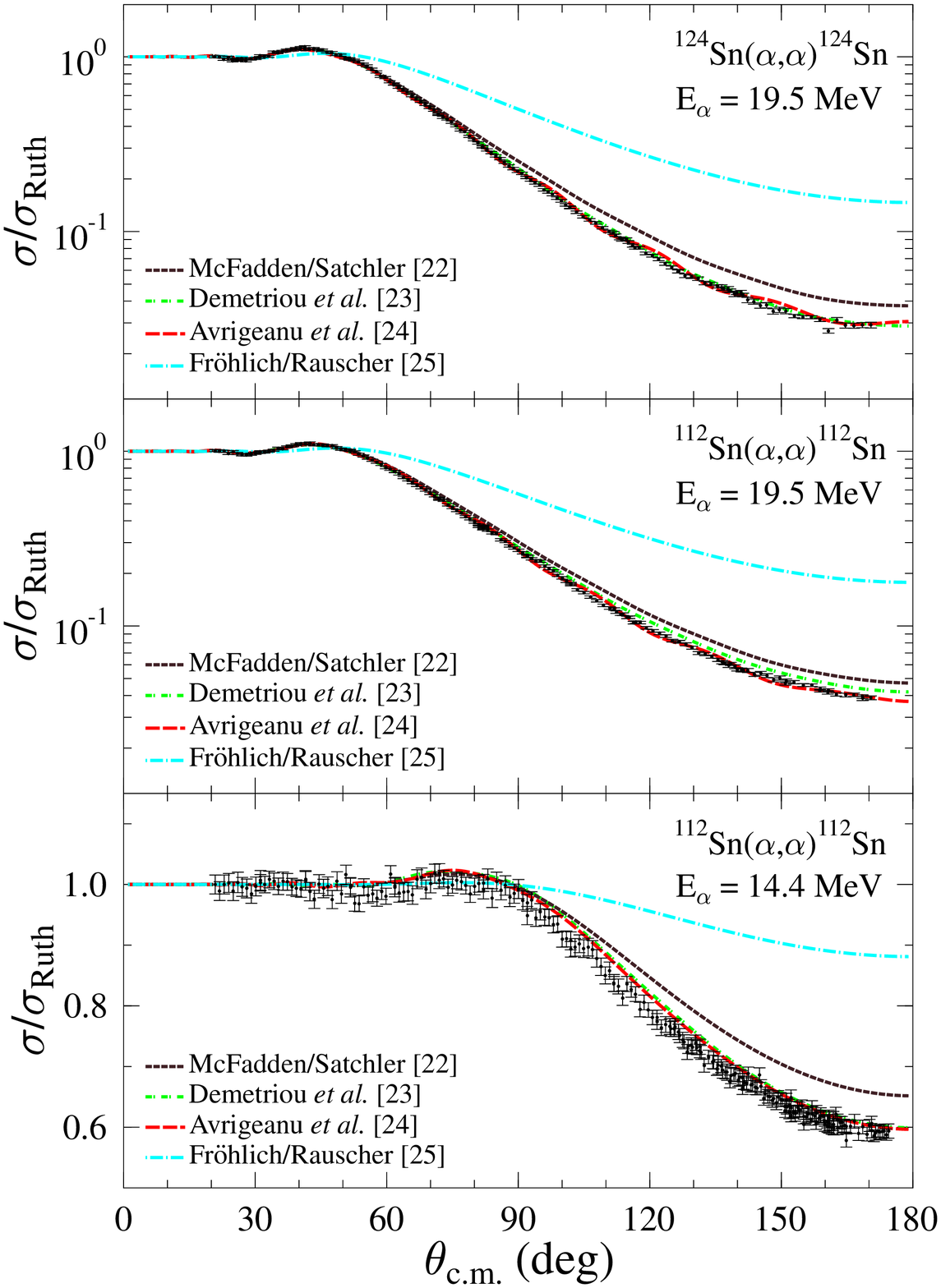}
\caption{
\label{fig:sigma}
(Color online) Ratio of the measured scattering cross sections of \Sna
\raa \Sna\ at 14.4\,MeV (lower diagram) and 19.5\,MeV (middle) and
\Snb \raa \Snb\ at 19.5\,MeV (upper) to the Rutherford cross
section. The predictions from the global \anucpots\ from \cite{McFa66}
(short dashed line), potential I from \cite{Deme02} (short dash-dotted
line), \cite{Avri03} (long dashed line) and \cite{Raus03} (long
dash-dotted line) are also shown. The overlap of the lines,
mainly at forward angles, complicates their distinction. Note
the logarithmic scale for the upper and middle diagrams
(E$_\alpha$\,=\,19.5\,MeV) and the linear scale for the lower diagram
(E$_\alpha$\,=\,14.4\,MeV).
}
\end{figure}

From the yield in the elastic peaks, the elastic scattering
cross section is calculated. The data is
normalized to the Rutherford cross section of point-like charged
particles. The experimental results are shown in
Fig.~\ref{fig:sigma}. Note that the measured cross
sections cover more than four orders of magnitude in the whole angular
range. Nevertheless, typical uncertainties remain below 3--4 \% for
all measured data points, including systematic and statistical
uncertainties. 

\section{Optical model analysis}
\label{sec:opt_mod}
\subsection{Angular distributions: comparison to theory}
\label{subsec:ang_theo}
The theoretical analysis of the angular distributions is performed
within the OM framework. The elastic scattering
cross section can be calculated from the Schr\"odinger equation with
the complex nuclear potential $U(r)$ given by
\begin{equation}
U(r) = V_{\rm{C}}(r) + V(r) + iW(r)
\label{eq:pot}
\end{equation}
with the Coulomb potential $V_{\rm{C}}(r)$, the real part $V(r)$, and
the imaginary part $W(r)$ of the nuclear potential. 

The calculated differential cross sections for four different global
\anucpots\ are also presented in Fig.~\ref{fig:sigma}. The
four--parameter Woods-Saxon potential from McFadden and Satchler
\cite{McFa66} provides a rough description of the experimental data,
overestimating the cross section in all three cases for backward
angles. Potential I from Demetriou {\it et al.}~\cite{Deme02} presents
a good reproduction of the measured angular distributions, with a
slight overestimation of the scattering cross sections at backward
angles for the reaction \Sna \raa \Sna\ at
E$_\alpha$\,=\,19.5\,MeV. This potential, based on a double folding
parameterization in its real part and a volume Woods-Saxon potential
in its imaginary part, provides a good description of previous \raa,
\ran\ and \rag\ cross section data.

The potential of Avrigeanu {\it et al.}~\cite{Avri03}, resulting from
the investigation of $\alpha$ scattering data at energies around the
Coulomb barrier for $A \approx$ 100 nuclei, is also compared to the
measured angular distributions. The potential is consistent with the
experimental data, although the corresponding cross section presents a
diffraction pattern at backward angles which is not observed in the
measured data. Finally, the potential presented by Fr\"ohlich and
Rauscher \cite{Raus03}, overestimates the cross section in all
cases. This potential is an expansion of the McFadden and Satchler's
potential to include \ran\ and \rag\ cross section data at energies
close to the astrophysically relevant region.
 
\begin{figure}[hbt]
\includegraphics[ width = 85 mm, clip]{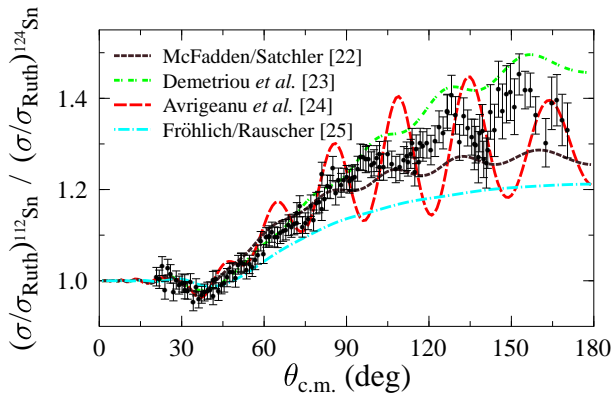}
\caption{
\label{fig:ratio}
(Color online) Ratio of the scattering cross sections
$(\sigma/\sigma_{\rm{Ruth}})_{\rm{^{112}Sn}}/(\sigma/\sigma_{\rm{Ruth}})_{\rm{^{124}Sn}}$
at E$_\alpha$\,=\,19.5\,MeV versus the angle in the center-of-mass
frame. The predictions of the considered global \anucpots\ are also
shown. Minor differences for the c.m. angle (below 0.2$^\circ$) in the
transformation to c.m. angles due to the different masses of \Sna\ and
\Snb\ are neglected.
}
\end{figure}

The depiction of the scattering cross section given by the different global
potentials (Fig.~\ref{fig:sigma}) makes very difficult to state which
potential provides the correct description of the experimental data. A
global \anucpot\ must be able to describe the scattering cross section
data along an isotopic chain, in order to demonstrate its reliability when
extrapolating to unstable nuclei of interest for astrophysical
applications. Despite the striking qualitative
similarities, the elastic scattering cross sections of \Sna\ and \Snb\
at E$_\alpha$\,=\,19.5\,MeV differ by roughly 30\% at very backward
angles. In Fig.~\ref{fig:ratio}, which shows the ratio of the two
cross sections (divided by the Rutherford cross section) as a function
of angle, all global \anucpots\ of~\cite{McFa66,Deme02,Avri03,Raus03}
fail to reproduce either the strength or the oscillation pattern for
backward angles. Thus, the use of these potentials in the extrapolation to
more proton--rich species (of interest in {\it p} process
nucleosynthesis) should be questioned. 

The following section studies the angular distributions, extracting a
local optical potential from the experimental data available.

\subsection{Angular distributions: present experiment}
\label{subsec:ang_pres}
In the present analysis, the real part of the potential is derived from a
double-folding procedure with two adjustable parameters:
\begin{equation}
V(r) = \lambda \cdot V_{\rm F}(r/w)
\end{equation}
where $V_{\rm{F}}(r)$ is the double-folding potential which is calculated
according to \cite{Kobo84,Satc79,Abel93,Atzr96} using the computer code
DFOLD. The required density distributions of the $\alpha$
particle and the $^{112,124}$Sn nuclei were derived from measured
charge density distributions \cite{Vrie87}. We vary the strength of the
double-folding potential by the parameter $\lambda$,
adopting values around 1.2 and 1.4 (similar to previous
works \cite{Atzr96,Mohr97,Fulo01}). This reduces the 
so-called {\it family problem} of \anucpots\ at low energies (see
Refs. \cite{Mohr97,Fulo01} for detailed discussion). The width of the
potential is adjusted using the parameter width $w$. We find values
close to 1 for our data.

The different optical potentials can be compared through their total
strengths or volume integrals, normalized to the number of interacting
nucleon pairs ($A_{\rm P} A_{\rm T}$), defined for both real and
imaginary parts of the nuclear potential:
%
\begin{eqnarray}
J_{\rm{R}} & = &
        \frac{1}{A_{\rm P} A_{\rm T}} \, \int V(r) \, d^3r
\label{eq:vol_r} \\
J_{\rm{I}} & = &
        \frac{1}{A_{\rm P} A_{\rm T}} \, \int W(r) \, d^3r
\label{eq:vol_i} 
\end{eqnarray}
Both volume integrals are negative. In this work, we will only
consider their absolute values.

The strength parameter $\lambda$ has been adopted to take the linear
form
\begin{eqnarray}
\lambda~=~\frac{a^* + b^* \cdot E_{\rm{c.m.}}}{J_{\rm{R,0}}}
\end{eqnarray}
where $J_0$ is the volume integral of the double folding
potential $V_{\rm F}(r/w)$. The values for the parameters
$a^*$, $b^*$ and $J_{\rm{R,0}}$, as extracted from the scattering
data, are listed in Table~\ref{tab:real_param}. The linear
energy dependence adopted for \Sna\ has been applied also for \Snb\ ,
although we have measurements for only one energy for this
nucleus. We check the validity of this linear dependence in
Sect.~\ref{subsec:ang_dist}, by analyzing the scattering data at
higher energies. The parameter $w$ allows a fine-tuning of the
potential width; it remains very close to unity. A significant
deviation of $w$ from unity for stable nuclei, where the neutron and
proton densities are very similar, would indicate that the
nucleon-nucleon interaction is not well chosen. However, for nuclei
with extremely high neutron-to-proton ratio one may expect the
formation of neutron skins; in this case, such a deviation of $w$ from
unity should be found for nucleon density distributions derived from
the proton density only.
\begin{table}
\caption{
Optical potential parameters for the real ($a^*$, $b^*$,
$J_{\rm{R,0}}$, $w$) and imaginary parts of the nuclear
potential. The parameters of the volume (V) and surface (S)
Woods-Saxon potentials ($W$, $R$, $a$) used in the imaginary part of
the nuclear potential are shown together with the volume integral
$J_{\rm{I}}$.
}
\vspace{0.5cm}
{
\begin{tabular}{c|cccc|ccccccc}
Isotope & $a^*$ & $b^*$ & $J_{\rm{R,0}}$ & $w$ & $W_{\rm{{V}}}$ &
        $R_{\rm{V}}$ & $a_{\rm{V}}$ & $W_{\rm{S}}$ & $R_{\rm{S}}$ &
        $a_{\rm{S}}$ & $J_{\rm{I}}$ \\ 
        & (MeV$\cdot$fm$^3$) & (fm$^3$) & (MeV$\cdot$fm$^3$) & & (MeV)
        & (fm) & (fm) & (MeV) & (fm) & (fm) & (MeV$\cdot$fm$^3$) \\
\hline\hline
\Sna & 352.92 & -0.652 & 277.85 & 1.004 & -3.137 & 1.737 & 0.341 &
        356.36 & 1.252 & 0.218 & 97.16 \\
\Snb & 355.12 & -0.652 & 274.90 & 1.006 & -2.467 & 1.723 & 0.296 &
        212.22 & 1.230 & 0.299 & 74.29 \\
\end{tabular}
}
\label{tab:real_param}
\end{table}

\begin{figure}[hbt]
\includegraphics[ width = 85 mm, clip]{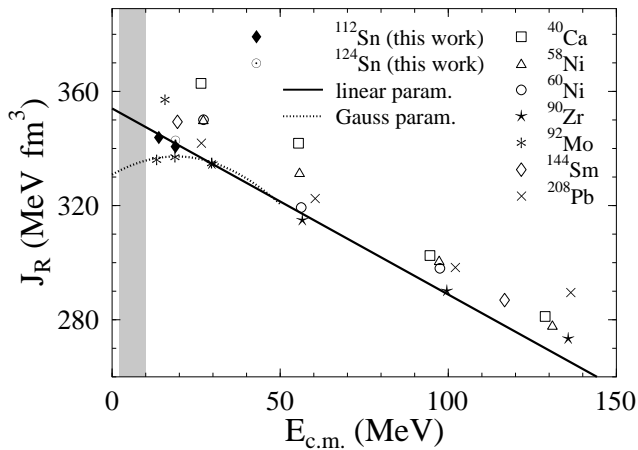}
\caption{
\label{fig:vol_int_jr}
Volume integral values $J_{\rm{R}}$ from the analysis of the
measured data, together with values obtained in the study of elastic
scattering data on doubly-magic and semi-magic nuclei
\cite{Atzr96,Mohr97,Fulo01}. The solid line shows a linear
parameterization of $J_{\rm{R}}$ extracted from the data. The Gauss
parameterization proposed by \cite{Deme02} for the description of low
energy data (dotted line) is also shown below 50 MeV. The
astrophysically relevant energy region for {\it p} process
nucleosynthesis is shown by the grey area.
}
\end{figure}

The volume integral $J_{\rm{R}}$ for the real part of the nuclear potential
adopted for the tin isotopes are shown in
Figure~\ref{fig:vol_int_jr}, together with the values derived from the
analysis of elastic scattering data on several doubly-magic and
semi-magic nuclei in the energy range up to 140 MeV
\cite{Atzr96,Mohr97,Fulo01}. The data show a systematic smooth linear
decrease at higher energies, in good agreement with the
parameterization adopted for the \Sna\ and \Snb\ isotopes (solid
line). The Gaussian parameterization, first suggested
by~\cite{Mohr00c} and modified in \cite{Fulo01,Deme02}, is shown as a
dotted line up to 50 MeV, which they proposed for astrophysical energies. 
 
The imaginary part of the nuclear potential has been chosen as a sum
of volume and surface Woods-Saxon potentials. The potential parameters
(potential depth $W_i$, radius $r_i=R_i \cdot A^{1/3}$ and
diffuseness $a_i$, with $i$=V,S) are listed in
Table~\ref{tab:real_param}. The relative weight between the volume and the
surface terms of the imaginary part of the nuclear potential is
$J_{\rm{I,V}}$ = 0.22 $\cdot J_{\rm{I,S}}$, as found in a study of the
elastic scattering data in the $A \approx$ 100 mass region
\cite{Gala04}. This dominance of the surface Woods-Saxon term at
energies close to the Coulomb barrier provides a better description of
$\alpha$ capture data at the astrophysically
interesting energy window \cite{Raus00b}.

\begin{figure}[hbt]
\includegraphics[ width = 85 mm, clip]{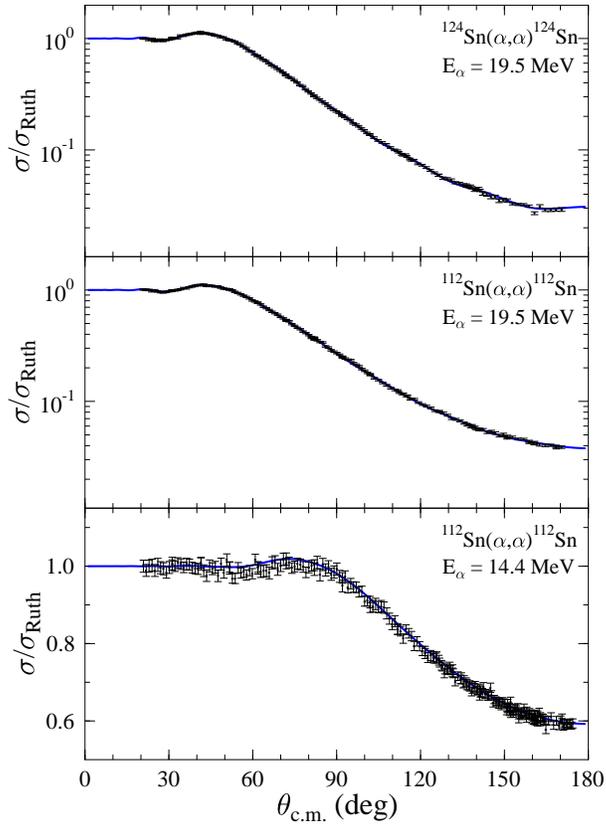}
\caption{
\label{fig:sigma_dgr}
Ratio of the measured scattering cross sections to the Rutherford cross
section (same as in Fig.~\ref{fig:sigma}) including the results of the
OM analysis performed for both tin nuclei (solid line).
}
\end{figure}

\begin{figure}[hbt]
\includegraphics[ width = 85 mm, clip]{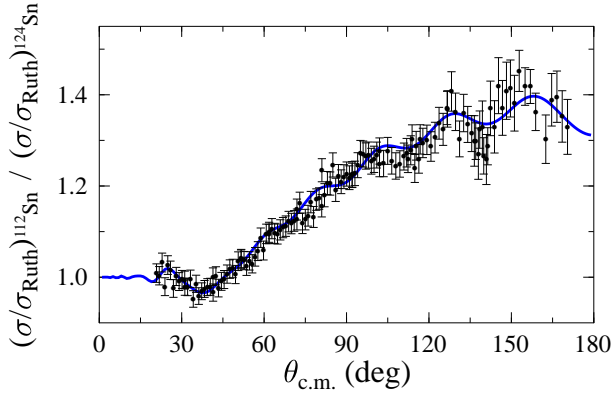}
\caption{
\label{fig:ratio_dgr}
Ratio of the scattering cross sections
$(\sigma/\sigma_{\rm{Ruth}})_{\rm{^{112}Sn}}/(\sigma/\sigma_{\rm{Ruth}})_{\rm{^{124}Sn}}$
at E$_\alpha$\,=\,19.5\,MeV versus the angle in the center-of-mass
frame including the results
of the OM analysis performed in both tin nuclei (solid line). A
significant improvement of the description of the experimental data
is achieved.
}
\end{figure}

The results of the OM analysis are compared in Fig.~\ref{fig:sigma_dgr}
to the experimental data from the three measured angular
distributions. An excellent agreement is observed. An average value of
$\chi^2_{red}$\,=\,1.1 results from the analysis. Unlike other
potentials, the ratio of the cross sections for the tin
isotopes is reproduced with higher accuracy, as shown in
Fig.~\ref{fig:ratio_dgr}.

In order to test the reliability of the obtained optical potentials,
and observe its possible energy dependence, the following section
compares the results of the OM analysis to other elastic scattering data
measured at higher energies.

%
\subsection{Angular distributions: comparison to other experiments}
\label{subsec:ang_dist}
\begin{figure}[hbt]
\includegraphics[ width = 105 mm, clip]{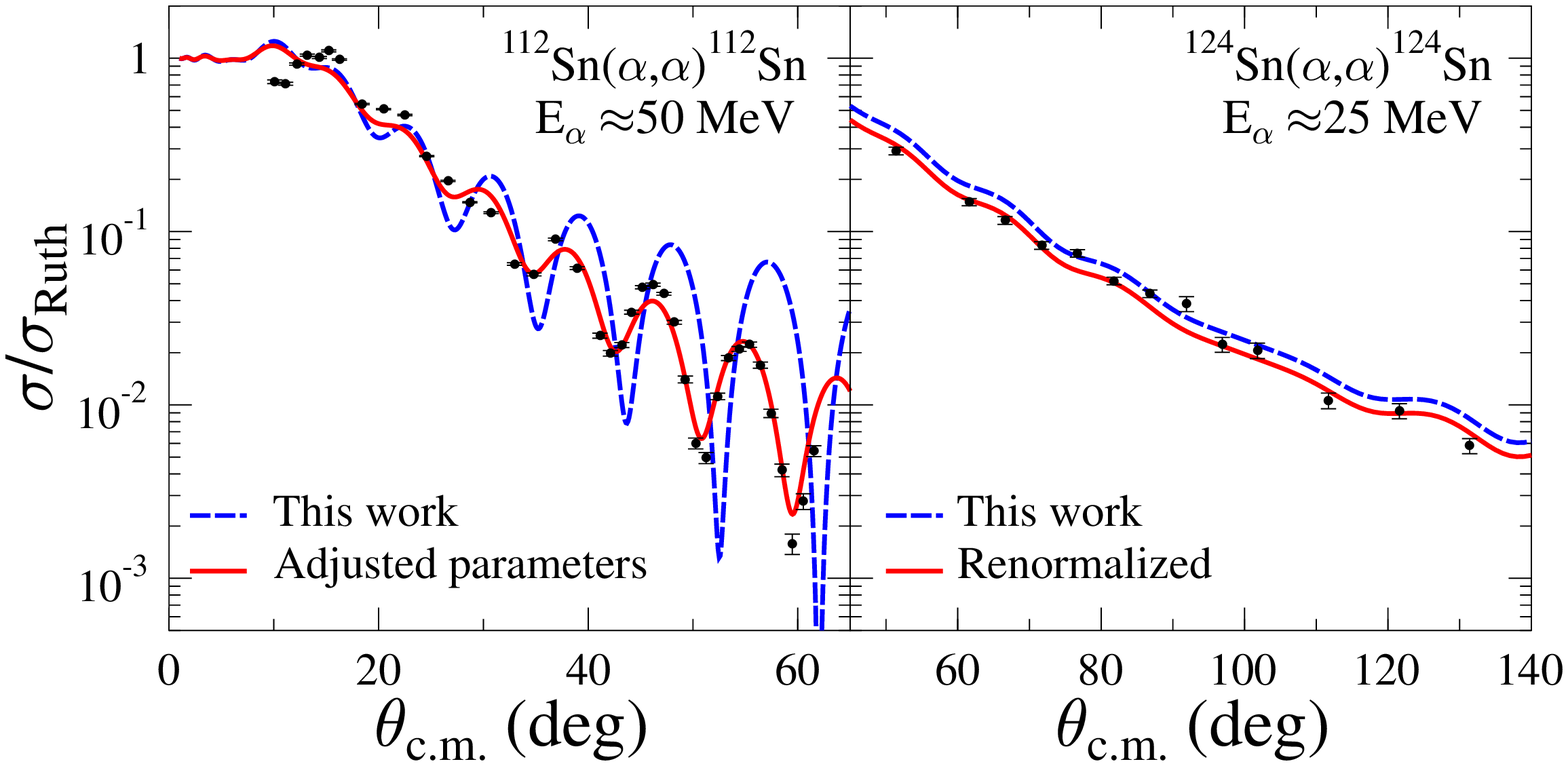}
\caption{
\label{fig:liter_aa}
(Color online) Elastic scattering data from the reactions \Sna \raa
\Sna\ at E$_\alpha \approx$ 50 MeV \cite{Burt90} (left part of the
Figure) and \Snb \raa \Snb\ at E$_\alpha \approx$ 25 MeV \cite{Besp92}
(right part of the Figure). The predictions from the derived
potentials from the elastic scattering data presented in this work are
also shown (dashed line). Results from a readjustment of the
potential parameters (left) and a normalization of the cross section
(right) are shown as a solid line. For details see text.
}
\end{figure}

The angular distribution of elastically scattered $\alpha$ particles
on the tin isotopes \Sna\ \cite{Burt90} and \Snb\ \cite{Besp92} has
been measured at energies far above the Coulomb barrier. These data
are shown in Fig.~\ref{fig:liter_aa} for the reactions
\Sna \raa \Sna\ at E$_\alpha \approx$ 50 MeV (left part of the
Figure) and \Snb \raa \Snb\ at E$_\alpha \approx$ 25 MeV (right part
of the Figure). In addition, the predictions from the optical
potentials derived from the analysis of our elastic scattering data
are shown.

The reproduction of the data from the reaction \Snb \raa \Snb\
\cite{Besp92} is satisfactory. The diffraction pattern shown by the
experimental data from the reaction \Sna \raa \Sna\ \cite{Burt90},
measured at energies far above from the Coulomb barrier, is not
described well. However, a minor variation of the potential
parameters (solid line) in which the contribution of the volume term to the
imaginary part of the nuclear potential is increased
($J_{\rm{I,V}}$= 0.79 $J_{\rm{I,S}}$) provides an improved description of
the scattering data from \cite{Burt90}. In the case of the scattering data
from \cite{Besp92}, closer to the energy region measured in this
work, a renormalization of the scattering data with a factor of 1.2
would provide a better agreement between the experimental data and the
calculated cross section.   

From this analysis, the imaginary part of the nuclear potential shows a
stronger energy dependence at energies higher than those considered in
this work. Without more experimental data for the nuclei
studied, it is not possible to predict the possible energy dependence
of both terms in the imaginary part of the nuclear potential.

\subsection{Excitation function: comparison to other experiments}
\label{subsec:exc_func}

\begin{figure}[hbt]
\includegraphics[ width = 85 mm, clip]{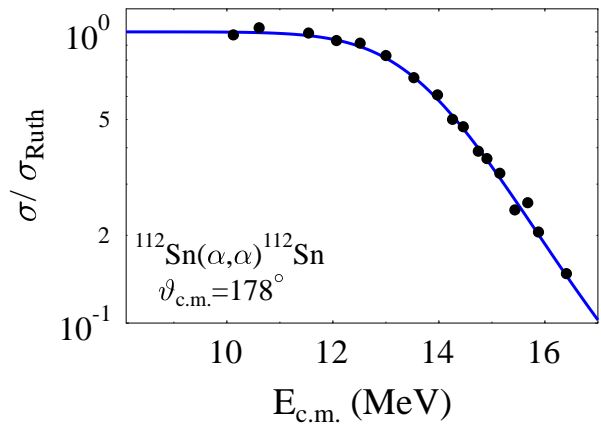}
\caption{
\label{fig:exc_funct}
Excitation function of the nucleus \Sna\ at $\vartheta_{\rm{c.m.}}$ =
178$^\circ$ \cite{Bada78}. The results from the derived \anucpot\
are also shown (solid line). The nice reproduction of the
experimental data confirms the good knowledge of the \anucpot\ in this
energy region. 
}
\end{figure}

The excitation function of the elastically scattered $\alpha$
particles on the nucleus \Sna\ at very backward angles
($\vartheta_{\rm{c.m.}}$ = 178$^\circ$) was measured by Badawy {\it et
  al.}~\cite{Bada78} at different energies below and above the Coulomb
barrier. The experimental data are shown in
Fig.~\ref{fig:exc_funct}. The successful reproduction of the
experimental data confirms the good knowledge of the \anucpot\ (solid
line) in the considered energy region.

\section{The \Sna \rag $^{116}$Te reaction}
\label{sec:sn_ag_te}

A relevant test of the potentials for astrophysical purposes consists
of the reproduction of the \rag\ reaction cross sections at energies close
to the Gamow window. In this section, the main features of the
statistical model are presented, followed by a comparison of existing
experimental data from the reaction \Sna \rag $^{116}$Te to the
predictions from the different \anucpots.

The main ingredients of the statistical model (Hauser-Feshbach
approach)~\cite{Haus52} in the calculation of reaction rates under
astrophysical conditions are transmission coefficients (particle and
radiative), nuclear level densities and optical potentials
~\cite{Raus00a}. These elements allow the calculation of the reaction
cross section in astrophysical scenarios. Once $\sigma^* (E)$ is
calculated, considering the case of the $\alpha$ capture
reaction, the reaction rate per particle pair at a given stellar
temperature $T^*$ is defined by \cite{Raus00a}:

\begin{eqnarray}
\label{eqn:rate}
\langle \sigma v \rangle^*~=~\biggl(\frac{8}{\pi\mu}\biggr)^{1/2}
\frac{1}{(kT^*)^{3/2}} \int_0^\infty \sigma_{(\alpha,\gamma)}^*(E)~
E ~ \exp \biggl(-\frac{E}{kT^*}\biggr) ~ dE
\end{eqnarray}
\noindent
by folding the stellar reaction cross section
$\sigma_{(\alpha,\gamma)}^*(E)$ with the Maxwell-Boltzmann velocity
distribution of the incident particles. $\mu$ is the reduced
mass of the system. In a stellar plasma, nuclei are in thermal
equilibrium with the environment and therefore can be found also in
excited states. The stellar reaction cross section
$\sigma^*=\sum_{\lambda \nu} \sigma^{\lambda \nu}$ includes
transitions from all populated target states $\lambda$ to all
energetically reachable final states $\nu$ whereas a laboratory cross
section $\sigma^\mathrm{lab}=\sum_\nu \sigma^{0 \nu}$ only accounts
for transitions from the ground state of the target. However, for the case of
$^{112}$Sn($\alpha$,$\gamma$) the stellar enhancement
$\sigma^*/\sigma^\mathrm{lab}$ is negligible in the {\it p} process
temperature range because no low-lying excited states in \Sna\ are
available for population. The product in the integrand of
Eq.~\ref{eqn:rate} leads to a maximum, defining an effective energy
window (the so-called {\it Gamow window}) where most of the reactions
occur. The reaction cross section should be determined in this
energy region. The photo-disintegration rate is then derived from the
capture rate applying detailed balance (see e.g. \cite{Raus00a}). 
 
During {\it p} process nucleosynthesis, typical temperatures of
2.0 $\leq T_9 \leq$ 3.0 are reached (where $T_9$ is the temperature in
GK). This corresponds to an energy
window between 5.1 and 9.6 MeV for the $\alpha$ capture reaction, or
between 4.2 and 8.7 MeV for the photon-induced \rga\ reaction
($Q_\alpha (^{116}$Te) = 930 keV)~\cite{Mohr03b}. 

\begin{figure}[hbt]
\includegraphics[ width = 85 mm, clip]{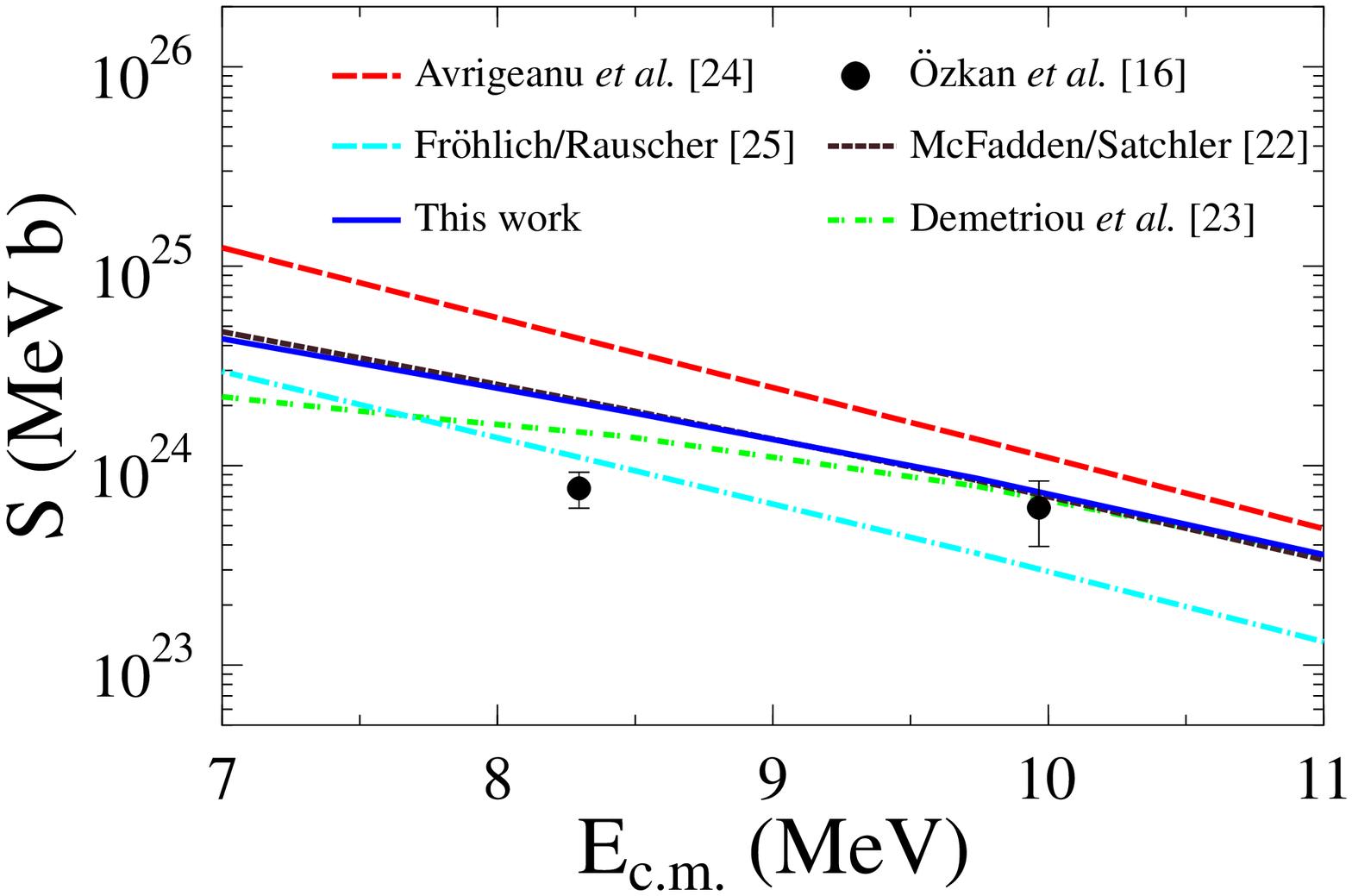}
\caption{
\label{fig:alpha_gamma}
(Color online) Astrophysical S-factor of the \Sna \rag $^{116}$Te capture
reaction. The experimental data from \cite{Ozka02} are compared
to the predictions from the global \anucpots\ and the optical potential
obtained from the analysis of the elastic scattering data. The
potential from \cite{Raus03} presents the best reproduction of the
experimental data. The prediction from the optical potential derived
in this work presents a satisfactory description of the measured cross
sections, similar to the potential from \cite{McFa66}.
}
\end{figure}

Due to the astrophysical interest, the laboratory \rag\ reaction cross
section on the nucleus \Sna\ has been measured \cite{Ozka02} at
energies close to the Gamow window. The astrophysical S-factor of the reaction
\Sna \rag$^{116}$Te is shown in Fig.~\ref{fig:alpha_gamma}. In
addition, the predictions from the global \anucpots\ and the optical
potential derived in this work are also plotted. Because of the scarce
experimental data available on the \Sna \rag$^{116}$Te reaction, we
can only perform a very limited comparison.

The global parameterization from \cite{Raus03} provides a satisfactory
description of the few experimental data, in contrast to its poor
agreement with the elastic scattering data. The other potentials, with
the exception of \cite{Avri03}, reproduce the cross section data
well. The potential obtained from the analysis of the elastic
scattering data presented in this work (Sec.~\ref{subsec:ang_pres})
provides a satisfactory description of the \rag\ data, very similar to
that of \cite{McFa66}. However, further $\alpha$ capture experiments on the
nucleus \Sna\ should be performed in order to cover the whole {\it
  Gamow window}. Experiments are under
way~\cite{Rapp}. These data should help to determine the experimental
energy dependence of the astrophysical S-factor.

As mentioned, once the astrophysical \rag\ capture cross section has
been calculated, it is possible to derive the corresponding
astrophysical capture and photo-disintegration
rates~\cite{Raus01,Raus00a}. For a comparison of the results provided
by the different parameterizations
of~\cite{McFa66,Deme02,Avri03,Raus03}, the variation of the obtained
reaction rates is shown in Table~\ref{tab:reac_rate}, where the ratios
of rates with respect to the rate obtained using \cite{Raus03} are
presented. All other ingredients of the 
statistical model calculations have been kept fixed as in \cite{Raus00a}. 
\begin{table}
\caption{
Ratio between the reaction rates obtained from the different \anucpots\
compared to the reaction rate of \cite{Raus03}, $\langle \sigma v
\rangle ^*_{[i]}$/$\langle \sigma v \rangle^*_{[25]}$, as a function
of the temperature $T_9$. The typical temperature range for {\it p}
process nucleosynthesis (2.0 $\leq T_9 \leq$ 3.0) is shown.
}
\vspace{0.5cm}
{
\begin{tabular}{c|cccc}
$T_9$ & \multicolumn{4}{c}{$\langle \sigma v \rangle ^*_{[i]} /
    \langle \sigma v \rangle^*_{[25]}$} \\

& \cite{McFa66} & \cite{Deme02} & \cite{Avri03} & This work \\
\hline \hline
\, 0.2 \, & \, 0.784 \, & \, 0.126 \, & \, 4.760 \, & \, 0.672 \, \\ 
\, 0.4 \, & \, 0.794 \, & \, 0.130 \, & \, 4.719 \, & \, 0.681 \, \\ 
\, 0.6 \, & \, 0.844 \, & \, 0.167 \, & \, 4.726 \, & \, 0.722 \, \\ 
\, 0.8 \, & \, 0.915 \, & \, 0.311 \, & \, 4.664 \, & \, 0.786 \, \\ 
\, 1.0 \, & \, 0.993 \, & \, 0.483 \, & \, 4.610 \, & \, 0.855 \, \\ 
\, 1.5 \, & \, 1.187 \, & \, 0.542 \, & \, 4.479 \, & \, 1.044 \, \\ 
\hline
\, 2.0 \, & \, 1.397 \, & \, 0.621 \, & \, 4.327 \, & \, 1.262 \, \\ 
\, 2.5 \, & \, 1.605 \, & \, 0.846 \, & \, 4.169 \, & \, 1.495 \, \\ 
\, 3.0 \, & \, 1.795 \, & \, 1.142 \, & \, 4.035 \, & \, 1.714 \, \\ 
\hline
\, 3.5 \, & \, 1.961 \, & \, 1.453 \, & \, 3.941 \, & \, 1.927 \, \\ 
\, 4.0 \, & \, 2.106 \, & \, 1.735 \, & \, 3.868 \, & \, 2.106 \, \\ 
\, 5.0 \, & \, 2.343 \, & \, 2.217 \, & \, 3.809 \, & \, 2.422 \, \\ 
\, 6.0 \, & \, 2.500 \, & \, 2.515 \, & \, 3.804 \, & \, 2.638 \, \\ 
\, 8.0 \, & \, 2.664 \, & \, 2.808 \, & \, 3.821 \, & \, 2.856 \, \\ 
\, 10.0 \, & \, 2.752 \, & \, 2.962 \, & \, 3.800 \, & \, 2.981 \, \\ 
\end{tabular}
}
\label{tab:reac_rate}
\end{table}

The different potentials predict reaction
rates which deviate up to a factor of 8. In the typical temperature
window for the {\it p} process, the reaction rates obtained by using
the optical potential derived from the elastic scattering predict in average a
rate which is around 50\% higher than that of \cite{Raus03}, remaining
very close to the values from McFadden and Satchler~\cite{McFa66}. 

\section{Summary}
\label{sec:summary}

We have measured the elastic scattering cross section of \Snc \raa
\Snc\ at energies $E_\alpha$\,=\,19.5\,MeV and 14.4\,MeV. The data have
been compared to various global \anucpots\ . The potentials from
\cite{Deme02} and \cite{Avri03} provide a satisfactory description of
the elastic scattering data, in contrast to the potentials from
\cite{McFa66} and \cite{Raus03}, which deviate considerably from the
measured angular distributions. None of the global potentials is able to
reproduce the ratio between the cross sections on the
proton- and neutron-rich tin isotopes. Consequently, any extrapolation
to \anucpots\ for unstable neutron--deficient nuclei on the {\it p}
process path remains uncertain. 

The present analysis performed within the OM framework provided a remarkable
reproduction of the measured angular distributions. It has been
used in the analysis of literature data at different energies
\cite{Burt90,Besp92,Bada78}. The results fit well with the systematic
behavior of $\alpha$--nucleus folding potentials. 

Most of the global \anucpots\ (with the exception of \cite{Avri03}),
as well as the potential obtained from the OM analysis, describe the
few existing \Sna \rag $^{116}$Te cross section data points well; however,
the energy dependence of the astrophysical S-factor is not well
determined from the theoretical predictions. The resulting
stellar rates for the \Sna \rag $^{116}$Te as well as $^{116}$Te\,\rga
\Sna\ reactions deviate in the energy region
considered for {\it p} process calculations by up to a factor of 8. 

The present data provide an excellent tool to test the behaviour of
global \anucpots . Additional elastic scattering experiments are
needed in the {\it p} process mass range at energies around the
Coulomb barrier.
 
\begin{acknowledgments}
We would like to thank the cyclotron team of ATOMKI for the excellent
beam during the experiment. Two of us (D.~G., P.~M.)  gratefully
acknowledge the kind hospitality at ATOMKI. We thank N.\ T.\
Burtebayev and O.\ Bespalova for providing the numerical data for
Refs.~\cite{Burt90} and \cite{Besp92}, respectively. We also thank M.\
Howard for reading the manuscript carefully. This work was
supported by DFG (SFB634 and FOR\,272/2-2) and OTKA (T034259, T042733,
F043408, D048283). T.\ R.\ is supported by the Swiss NSF (grants
2024-067428.01, 2000-061031.02, 2000-105328). Zs.\ F.\ and Gy.\ Gy.\
acknowledge support from a Bolyai grant. 
\end{acknowledgments}

\end{document}